\documentclass[pra,twocolumn,showpacs]{revtex4}
\usepackage{amsmath,graphicx}
\begin{document}

\title{Geometric Quantum Computation on Solid-State Qubits}%
\author{Mahn-Soo Choi}%
\affiliation{Department of Physics, Korea University, Seoul 136-701,
  Korea}

\begin{abstract}
Geometric quantum computation is a scheme to use non-Abelian Holonomic
operations rather than the conventional dynamic operations to manipulate
quantum states for quantum information processing.  Here we propose a
geometric quantum computation scheme which can be realized with current
technology on nanoscale Josephson-junction networks, known as a
promising candidate for solid-state quantum computer.
\end{abstract}


\maketitle

\newcommand\ket[1]{|#1\rangle}%
\newcommand\bra[1]{\langle#1|}%
\newcommand\varE{\mathcal{E}}%
\newcommand\half{\frac{1}{2}}%

\section{Introduction}

The elementary units of quantum information processing are quantum
two-state systems, called quantum bits or ``qubits''. Unlike a classical
bit, a qubit can be in any superposition
$\alpha\ket{\bar0}+\beta\ket{\bar1}$ (with $\alpha$ and $\beta$
arbitrary complex numbers satisfying the normalization condition
$|\alpha|^2+|\beta|^2=1$) of the computational basis states
$\ket{\bar0}$ and $\ket{\bar1}$.  A qubit needs not only to preserve
quantum coherence for a sufficiently long time but also to allow an
adequate degree of controllability.  Among a number of ideas proposed so
far to realize qubits, the ones based on solid-state devices have
attracted interest due to the scalability for massive information
processing, which can make a quantum computer of practical
value~\cite{Averin01a}.

Another crucial element of quantum information processing is the ability
to perform quantum operations on qubits in a controllable way and with
sufficient accuracy.  In most proposed schemes such quantum operations
are unitary, and conventionally have been achieved based on quantum
dynamics governed by the Schr\"odinger equation.

Recently, it has been proposed that controllable quantum operations can
be achieved by a novel geometric principle as
well~\cite{Zanardi99a,Pachos00a}.  When a quantum system undergoes an
adiabatic cyclic evolution, it acquires a non-trivial geometric
operation called a holonomy.  Holonomy is determined entirely by the
geometry of the cyclic path in the parameter space, independent of any
detail of the dynamics.  If the eigenspace of the Hamiltonian in
question is non-degenerate, the holonomy reduces to a simple phase
factor, a Berry phase.  Otherwise, it becomes a non-Abelian unitary
operation, i.e., a non-trivial rotation in the eigenspace.  It has been
shown that universal quantum computation is possible by means of
holonomies only~\cite{Zanardi99a,Pachos00a}.  Further, holonomic quantum
computation schemes have intrinsic tolerance to certain types of
computational errors~\cite{Preskill99a,Ellinas01a}.

A critical requirements for holonomies is that the eigenspace should be
preserved throughout the adiabatic change of parameters, which is
typically fulfilled by symmetry~\cite{Wilczek84a}.  It is non-trivial to
devise a physical system with a proper eigenspace which will serve as a
computational space.  Recently a scheme for geometric quantum
computation with nuclear magnetic resonance was proposed and
demonstrated experimentally~\cite{Jones00a}.  A similar scheme was
proposed on superconducting nanocircuits~\cite{Falci00a}.  In these
schemes, however, geometrically available was only Abelian Berry phase,
and additional dynamic manipulations were required for universal quantum
computation.  A scheme based solely on holonomies has been proposed for
quantum optical systems~\cite{Pachos00b}.  However, it relies on
nonlinear optics, which may make this scheme less practical.  More
recently, another holonomic quantum computation scheme has been proposed
for trapped ions~\cite{DuanLM01a}.  Supposedly, it is the only
experimentally feasible scheme proposed so far for holonomic
computation.  Here we propose a scheme for holonomic quantum computation
on nanoscale Josephson networks, known as a promising candidate for
solid-state quantum computer~\cite{Makhlin99a,Nakamura99a,Mooij99a}.  It
relies on tunable Josephson junction and capacitive coupling, which are
already viable with current technology.

\section{Josephson Charge Qubits}

\begin{figure}
\begin{center}
\includegraphics{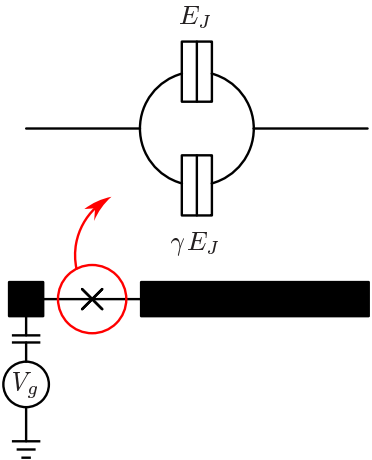}
\caption {A schematic of a ``Josephson charge qubit''. The cross denotes a
  \emph{tunable} Josephson junction, which consists of two parallel
  tunnel junctions (upper panel).  The strength of the effective
  Josephson coupling is tuned by flux threading the loop.  The gate
  voltage $V_g$ controls the induced charge on the box.}
\label{Holonomic:fig1}
\end{center}
\end{figure}

A ``Josephson (charge) qubit''~\cite{Makhlin99a,Nakamura99a} can be
realized by a small superconducting grain (a Cooper-pair box) of size
$\sim100\mathrm{nm}$, coupled to a large superconducting charge
reservoir or another Cooper-pair box via a Josephson junction, see
Fig.~\ref{Holonomic:fig1}.  The computational bases are encoded in two
consecutive charge states $\ket{\bar0}\to\ket{0}$ and
$\ket{\bar1}\to\ket{1}$ with $\ket{n}$ denoting a state with $n$
\emph{excess} Cooper pairs on the box.  States with more (or less)
Cooper pairs are suppressed due to the strong Coulomb repulsion (the
gate-induced charge $2en_g$ is tuned close to $1/2 \mod 1$),
characterized by the charging energy $E_C=(2e)^2/2C$ (with $C$ the total
capacitance of the box).  Excitation of quasiparticles is also ignored
assuming sufficiently low temperature.  The tunneling of Cooper pairs
across the junction, characterized by the Josephson coupling energy
$E_J$ ($\ll{}E_C$), allows coherent superposition of the charge states
$\ket0$ and $\ket1$.  A \emph{tunable} junction is attained by two
parallel junctions making up a SQUID (superconducting quantum
interference device) with a magnetic flux $\Phi$ threading through the
loop, see upper panel of Fig.~\ref{Holonomic:fig1}~\cite{Tinkham96a}.
Namely, in the two-state approximation, the Hamiltonian is written in
terms of the Pauli matrices $\sigma^z$ and
\begin{math}
\sigma^\pm = \frac{1}{2}\left(\sigma^x \pm i\,\sigma^y\right)
\end{math}
as~\cite{Falci00a,Makhlin99a}
\begin{equation}
\label{Holonomic:Junction}
H
= - \frac{1}{2}h(n_g)\sigma^z
- \frac{1}{2}\left[J(\phi)\sigma^+ + J^*(\phi)\sigma^-\right]
\end{equation}
where $h(n_g)=E_C(2n_g-1)$, $J(\phi)$ is the effective Josephson
coupling of the tunable junction (i.e. SQUID), and $\phi=\pi\Phi/\Phi_0$
(we assume that $|\phi|\leq\pi/2$) with $\Phi_0=hc/2e$ the
superconducting flux quantum.  Given Josephson energies $E_J$ and
$\gamma{}E_J$ ($\gamma>0$) of the two parallel junctions on a SQUID
loop, the magnetic flux gives rise to a phase shift $\alpha(\phi)$ as
well as an amplitude modulation $A(\phi)$ of the effective Josephson
coupling
\begin{math}
J(\phi) = 2E_J\, A(\phi) e^{-i\alpha(\phi)}
\end{math}.
$A(\phi)$ and $\alpha(\phi)$ are given by~\cite{Falci00a}
\begin{equation}
\label{Holonomic:JA}
A(\phi) = \sqrt{(1-\gamma)^2/4 + \gamma\cos^2\phi}
\end{equation}
and
\begin{equation}
\label{Holonomic:Jalpha}
\tan\alpha(\phi) = \frac{1-\gamma}{1+\gamma}\tan\phi \,,
\end{equation}
respectively.  It is worth noticing that for identical junctions
($\gamma=1$), (i) there is no phase modulation [$\alpha(\phi)=0$] and
(ii) the effective Josephson coupling can be turned off completely
[$J(\phi)=0$] at $\phi=\pi/2$.  In what follows, some tunable junctions
have $\gamma=1$ and others $\gamma\neq1$ depending on their roles for
the system.

Below we will demonstrate that one can obtain the three unitary
operations,
\begin{math}
U_Z(\varphi) = \exp\left(i\varphi\ket{\bar1}\bra{\bar1}\right)
\end{math}
(phase shift),
\begin{math}
U_X(\varphi) = \exp\left(i\varphi\sigma^x\right)
\end{math}
(rotation around $x$ axis), and
\begin{math}
U_{CZ}(\varphi)
=\exp\left(+i\varphi\ket{\bar1\bar0}\bra{\bar1\bar0}\right)
\end{math}
(controlled phase shift), on an arbitrary qubit or pair of qubits, using
\emph{geometric} manipulations only.  It is known that these unitary
operations form a universal set of gate operations for quantum
computation~\cite{Lloyd95a,DuanLM01a}.  Since the charge degrees of
freedom is used in the present scheme, the state preparation and the
state readout, which are another important procedures required for
quantum computation, can be achieved using the same methods used in
dynamical schemes~\cite{Makhlin99a}.

\section{Elementary Gates}

\begin{figure}
\begin{center}
\includegraphics{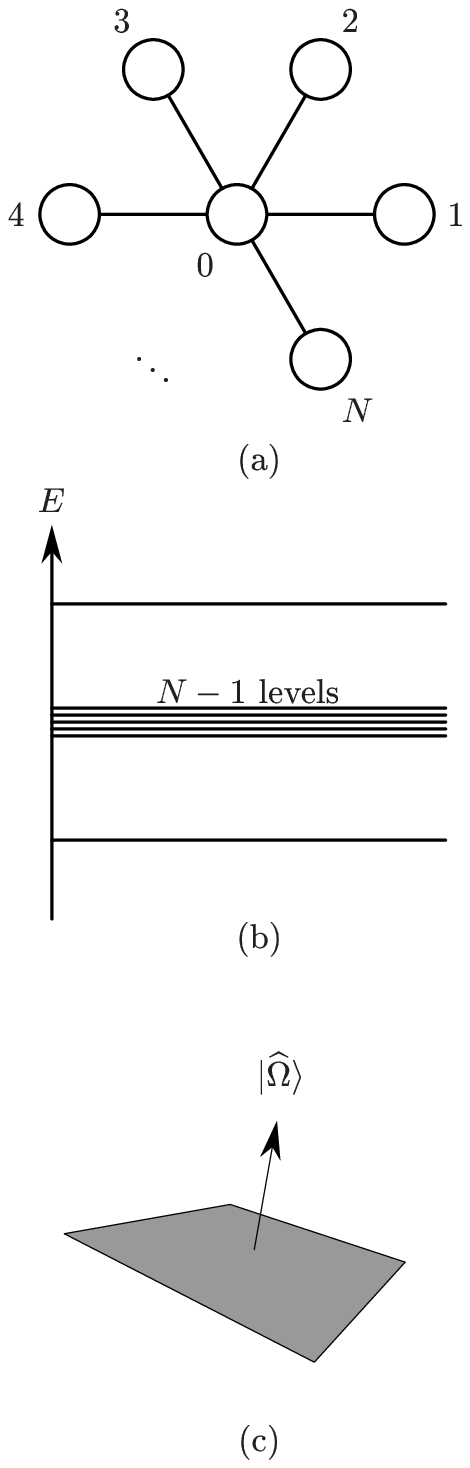}
\caption{(a) A schematic representation of the prototype model,
  Eq.~(\ref{Holonomic:Prototype}), which has a proper degenerate
  eigenspace for geometric quantum computation.  (b) The degenerate
  level structure of the model.  (c) The structure of the corresponding
  Hilbert space, in which the degenerate subspace is always
  perpendicular to $\ket\Omega$.}
\label{Holonomic:fig2}
\end{center}
\end{figure}

Before demonstrating the geometric implementation of elementary gates,
we suggest a prototype Hamiltonian which reveals the proper symmetry for
geometric manipulations in question.  All the Hamiltonians considered in
this paper share the following common structure:
\begin{equation}
\label{Holonomic:Prototype}
H
= \epsilon\ket{\hat0}\bra{\hat0}
- \frac{1}{2}\sum_{i=1}^N\left(
  \Omega_i\ket{\hat i}\bra{\hat0}
  + \Omega_i^*\ket{\hat0}\bra{\hat i}
\right) \,.
\end{equation}
Here $\Omega_i$ is the transition amplitude from the state $\ket{\hat0}$
to $\ket{\hat i}$ and $\epsilon$ is the energy of the state
$\ket{\hat0}$ measured from the degenerate energy of the states
$\ket{\hat i}$ ($i=1,\ldots,N$).  For our consideration below, one may
regard the state vector $\ket{\hat i}$ ($i=0,1,\ldots,N$) represent,
e.g., an excess Cooper pair on the $i$th superconducting box, see
Fig.~\ref{Holonomic:fig2} (a).  $\ket{\hat i}$ may also represent
electronic levels in atoms such as discussed in
Ref.~\onlinecite{DuanLM01a}.

As one changes the parameters $\Omega_i$, the Hamiltonian in
Eq.~(\ref{Holonomic:Prototype}) preserves the $(N{-}1)$-dimensional
degenerate subspace.  This can be clearly seen by defining
\begin{math}
\ket{\widehat\Omega} \equiv \Omega^{-1}\sum_{i=1}^N\Omega_i\ket{\hat i}
\end{math}
with
\begin{math}
\Omega^2 \equiv \sqrt{|\Omega_1|^2+\cdots+|\Omega_N|^2}
\end{math},
and rewriting $H$ as
\begin{equation}
\label{Holonomic:H0a}
H
= \epsilon\ket{\hat0}\bra{\hat0}
- \frac{1}{2}\Omega\left(
  \ket{\widehat\Omega}\bra{\hat0} + \ket{\hat0}\ket{\widehat\Omega}
\right) \,.
\end{equation}
The Hamiltonian in Eq.~(\ref{Holonomic:H0a}) corresponds to a particle
in a (biased) double-well potential in the tight-binding approximation.
Therefore, it has two eigenstates
\begin{math}
\ket{\lambda_\pm} = \frac{1}{\sqrt
  2}(\ket{\widehat\Omega}\mp\ket{\hat0})
\end{math}
with energies
\begin{math}
\lambda_\pm = \frac{1}{2}(\epsilon \pm \sqrt{\Omega^2+\epsilon^2})
\end{math}.
The other $N-1$ levels out of $N+1$ form a degenerate subspace
$\varE_{N-1}$ with energy $0$, which we will use later for a
computational subspace, see Fig.~\ref{Holonomic:fig2} (b).  Notice that
the degenerate eigenspace $\varE_{N-1}$ is always perpendicular both to
$\ket{\hat0}$ and $\ket{\widehat\Omega}$; as the parameters $\Omega_i$
change, $\ket{\widehat\Omega}$ rotates in the Hilbert space, and the
eigenspace $\varE_{N-1}$ is attached \emph{rigidly}, perpendicular to
$\ket{\widehat\Omega}$, see Fig.~\ref{Holonomic:fig2} (c).


\begin{figure}
\begin{center}
\includegraphics{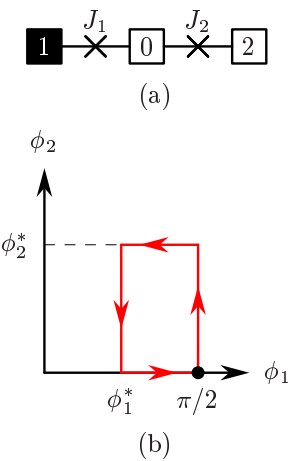}
\caption {Configuration for the phase-shift operation $U_Z$.
  Throughout the work, computational bases are encoded in the solid
  boxes and empty boxes serve as ancilla qubits.}
\label{Holonomic:fig3}
\end{center}
\end{figure}

We first show how to get the unitary operation $U_Z$ geometrically.  We
consider three Cooper-pair boxes coupled in series as shown in
Fig.~\ref{Holonomic:fig3}.  The first (solid) box encodes the
computational bases while the second and third (empty) boxes serve as
``ancilla'' qubits.  The Hamiltonian is given by [see
Eq.~\eqref{Holonomic:Junction}]
\begin{equation}
\label{Holonomic:H3}
H
= -\frac{1}{2}\sum_{n=1}^2\left(J_n\sigma_n^+\sigma_0^- + h.c.\right)
- \frac{1}{2}h\,\sigma_0^z
\end{equation}
with $h.c.$ meaning the Hermitian conjugate.  Comparing
Eq.~(\eqref{Holonomic:H3}) with the prototype Hamiltonian in
Eq.~(\ref{Holonomic:Prototype}), noticed are following correspondences:
\begin{math}
\ket{\hat 0} \to \ket{100}_{012}
\end{math}
($\ket{100}_{123}$ is short for $\ket{1}_0\ket0_1\ket0_2$),
\begin{math}
\ket{\hat 1} \to \ket{010}_{012}
\end{math},
\begin{math}
\ket{\hat 2} \to \ket{001}_{012}
\end{math},
\begin{math}
\ket{\hat 3} \to \ket{000}_{012}
\end{math},
$\Omega_1\to J_1$, $\Omega_2\to J_2$, and $\Omega_3=0$.  From this [or
direct diagonalization of the Hamiltonian Eq.~(\ref{Holonomic:H3}), of
course], one can see that the two states
\begin{equation}
\label{Holonomic:Ket1a}
\ket{\lambda_1} = J_2^*\ket{010}_{012} - J_1^*\ket{001}_{012}
\end{equation}
(not normalized) and
\begin{equation}
\label{Holonomic:Ket1b}
\ket{\lambda_2} = \ket{000}_{012}
\end{equation}
form a degenerate subspace with energy $-h/2$, which is preserved during
the change of $J_1$ and $J_2$ (equivalently $\phi_1$ and $\phi_2$).
Since the computational basis is only encoded in the ``true'' qubit (box
1), the total wave function $\ket\Psi$ of the logical block should be
initially prepared in a separable state with respect to the true qubit
and the ancilla qubits,
\begin{math}
\ket\Psi = \ket{\psi}_1\otimes\ket{\psi'}_{02}
\end{math}.
In other words, one should be able to turn off at will the \emph{tunable
  junction} 1, which should therefore be made of identical parallel
junctions ($\gamma_1=1$), see Eqs.~(\ref{Holonomic:JA}) and
(\ref{Holonomic:Jalpha}).  After a cyclic evolution of the parameters
$\phi_1$ and $\phi_2$ along a closed loop starting and ending at the
point with $\phi_1=\pi/2$ (i.e., $J_1=0$), the state $\ket{\lambda_1}$
acquires the Berry phase~\cite{Berry84a}.  For example, along the loop
depicted in Fig.~\ref{Holonomic:fig3} (b), the Berry phase is given by
\begin{equation}
\label{Holonomic:Berry1}
\varphi_B
= \frac{1-\gamma_2^2}{4}\int_0^{\phi_2^*}{d\phi_2}\;\left[
  \frac{1}{\cos^2\phi_1^*+A_2^2(\phi_2)}
  - \frac{1}{A_2^2(\phi_2)}
  \right] \,.
\end{equation}
(The Berry phase vanishes if $\gamma_2=1$ as expected~\cite{Berry84a}.)
The state $\ket{\lambda_2}$ remains unchanged.  Therefore, the cyclic
evolution amounts to $U_Z(\varphi_B)$.

Here it should be emphasized that although used is the \emph{Abelian}
Berry phase, the degenerate structure is crucial.  The dynamic phases
acquired by $\ket{\lambda_1}$ and $\ket{\lambda_2}$ are the same and
result in a trivial global phase.  In recently proposed
schemes~\cite{Jones00a,Falci00a}, which have no degenerate structure, at
least one dynamic manipulation was unavoidable to remove the dynamically
accumulated phases.  Another point to be stressed is that the phase
shift $\alpha_2(\phi_2)$ in the effective Josephson coupling
$J_2(\phi_2)$ is indispensable for the Berry phase.


\begin{figure}
\begin{center}
\includegraphics{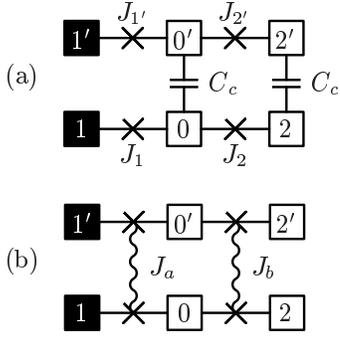}
\caption {(a) Configuration for the two-qubit gate $U_{CZ}$. (b)
  Effective ``joint-tunnel'' coupling (crossed wiggly lines) resulting
  from the capacitive coupling in (a).}
\label{Holonomic:fig4}
\end{center}
\end{figure}

The two-qubit gate operation $U_{CZ}$ can be realized geometrically
using capacitive coupling.  As shown in Fig.~\ref{Holonomic:fig4}~(a),
the ancilla qubits on different three-box systems are coupled in
parallel via capacitors with capacitance $C_c$.  It is
known~\cite{ChoiMS98f,Shimada00a,Averin91a,Matters97a} that for $C_c$
sufficiently larger than the self-capacitance $C$ of each box, the
states $\ket{01}_{00'}$ and $\ket{10}_{00'}$ are strongly favored over
the states $\ket{00}_{00'}$ and $\ket{11}_{00'}$, and the same for boxes
$2$ and $2'$ (recall that $n_g\approx{}1/2$ for each box).  This
effectively leads to a ``joint tunneling'', see
Fig.~\ref{Holonomic:fig4}~(b): Tunneling of a charge from box $1$ to $0$
should be accompanied by tunneling of another charge from $0'$ to $1'$.
The joint-tunneling amplitudes are given by
$J_a\approx{}4J_1J_{1'}^*/E_C$ and $J_b\approx{}4J_2J_{2'}^*/E_C$
\cite{ChoiMS98f}.  Then the total Hamiltonian has the form
\begin{multline}
\label{Holonomic:H4}
H = -\half\left[J_a\left(\sigma_1^+\sigma_0^-\right)
  \left(\sigma_{1'}^-\sigma_{0'}^+\right) + h.c.\right] \\
-\half\left[J_b\left(\sigma_2^+\sigma_0^-\right)
  \left(\sigma_{2'}^-\sigma_{0'}^+\right) + h.c.\right] - \half
h\left[\sigma_0^z-\sigma_{0'}^z\right] \,.
\end{multline}
In analogy with Eqs.~(\ref{Holonomic:Prototype}) and
(\ref{Holonomic:H3}), the above Hamiltonian has an eigenspace containing
the four degenerate states
\begin{equation}
\ket{\lambda_{00}} = \ket{00}_{11'}\otimes\ket{0101}_{00'22'} \,,
\end{equation}                             
\begin{equation}                           
\ket{\lambda_{01}} = \ket{01}_{11'}\otimes\ket{0101}_{00'22'} \,,
\end{equation}                             
\begin{equation}                           
\ket{\lambda_{11}} = \ket{11}_{11'}\otimes\ket{0101}_{00'22'} \,,
\end{equation} and
\begin{equation}
\ket{\lambda_{10}}
= J_b^*\ket{{10}}_{11'}\otimes\ket{0101}_{00'22'}
- J_a^*\ket{01}_{11'}\otimes\ket{0110}_{00'22'}
\end{equation} (not normalized)  with energy $-h$.
As in the previous case [see discussions below
Eq.~(\ref{Holonomic:Ket1b})], it is assumed that the tunable junctions
$J_1$ and $J_{1'}$ are composed of identical parallel junctions
($\gamma_1=\gamma_{1'}=1$), while $\gamma_2\neq1$ and
$\gamma_{2'}\neq1$.  One can fix $\phi_{1'}=\phi_{2'}=0$ and change the
parameters $\phi_1$ and $\phi_2$ along a closed loop starting from the
point with $\phi_1=\pi/2$ ($J_1=J_a=0$).  Upon this cyclic evolution,
the state $\ket{\lambda_{10}}$ acquires the Berry phase in
Eq.~\eqref{Holonomic:Berry1} while the other three states remain
unchanged, leading to the two-qubit gate operation $U_{CZ}(\varphi_B)$.
We mention that in this implementation of $U_{CZ}$, the capacitive
coupling is merely an example and can be replaced by any other coupling
that effectively results in a sufficiently strong Ising-type interaction
of the form $J_{ij}\sigma_i^z\sigma_j^z$.


\begin{figure}
\begin{center}
\includegraphics{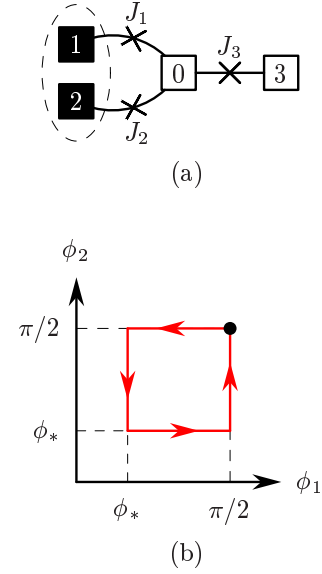}
\caption {(a) Configuration for a rotation around $x$ axis, $U_X$.  The
  computational bases is encoded over the two (solid) boxes, not in a
  single box.  (b) An adiabatic path in the parameter space to achieve
  $U_X$ ($\phi_3$ is kept constant all along the path).}
\label{Holonomic:fig5}
\end{center}
\end{figure}

In the demonstrations of implementing $U_Z$ and $U_{CZ}$ above, we have
encoded the bases $\ket{\bar0}$ and $\ket{\bar1}$ in a single
Cooper-pair box for simplicity.  To realize $U_X$, we need to encode the
basis states over two Cooper-pair boxes, e.g., box $1$ and $2$ in
Fig.~\ref{Holonomic:fig5}~(a): $\ket{\bar0}\to\ket{01}_{12}$ and
$\ket{\bar1}\to\ket{10}_{12}$.  It is straightforward to generalize the
above implementations of $U_Z$ and $U_{CZ}$ in this two-box encoding
scheme.  Now we turn to the remaining single-qubit gate operation $U_X$.
The Hamiltonian is given by
\begin{equation}
\label{Holonomic:H5}
H
= - \half\sum_{n=1,2,3}\left(J_n\,\sigma_n^+\sigma_0^- + h.c.\right)
  - \half h\sigma_0^z \,.
\end{equation}
The degenerate subspace is defined by the two eigenstates
\begin{equation}
\ket{\lambda_1}
= \left(J_2^*\,\ket{{10}}_{12} - J_1^*\,\ket{{01}}_{12}\right)\otimes
  \ket{00}_{03}
\end{equation}
(not normalized) and
\begin{multline}
\ket{\lambda_2} = \frac{J_3^*}{|J_1|^2+|J_2|^2}\left( J_1\ket{{10}}_{12}
  + J_2\ket{{01}}_{12}
\right)\otimes\ket{00}_{03} \\
- \ket{00}_{12}\otimes\ket{01}_{03}
\end{multline}
(not normalized) both with energy $-h/2$.  In this case, it is required
that $\gamma_1=\gamma_2=1$ but $\gamma_3\neq1$ [see discussions below
Eq.~(\ref{Holonomic:Ket1b})].  As an example, we take a closed loop
shown in Fig.~\ref{Holonomic:fig5}~(b) (one may choose any path starting
and ending at $\phi_1=\phi_2=\pi/2$, i.e., $J_1=J_2=0$) with $\phi_3$
fixed.  The adiabatic theory of holonomies~\cite{Wilczek84a} ensures
that from this adiabatic cycle, a state $\ket\psi$ initially belonging
to the eigenspace undergoes a change to $\ket{\psi'}=U\ket\psi$.  The
unitary operator $U$ is given by
\begin{math}
U = U_Z^\dag(\varphi')U_X(\varphi)U_Z(\varphi')
\end{math}
with $\varphi'=\alpha_3(\phi_3)/2-\pi/4$ and
\begin{equation}
\varphi
= 2\int_{\cos\phi_*}^1
  \frac{dx\;\cos\phi_*}{%
    (x^2+\cos^2\phi_*)
    \sqrt{1+(x^2+\cos^2\phi_*)/A_3^2(\phi_3)}} \,.
\end{equation}
Removing the first and last factors of $U_Z$ (if necessary) with
additional phase-shift operations, one can achieve $U_X$.

\begin{figure}
\begin{center}
\includegraphics{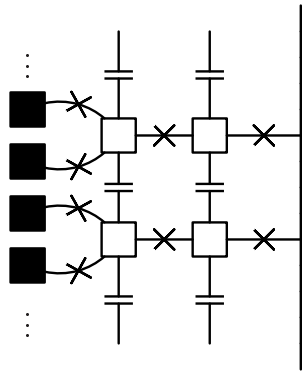}
\caption {A computational network for universal holonomic quantum
  computation.}
\label{Holonomic:fig6}
\end{center}
\end{figure}

Finally, a quantum network can be constructed as in
Fig.~\ref{Holonomic:fig6} to perform all the unitary operations
discussed above (and hence universal quantum computation) geometrically.
It is noted that the coupling capacitance $C_c$ is not tunable, but it
suffices to have a control over each tunable junction (i.e. SQUID) and
gate voltage.

\section{Discussion}

So far we have discussed fundamental requirements for geometric
manipulations in idealized systems.  In this section, we discuss several
situations one may come up with when attempting an experimental
realization of the present scheme.

First of all, the present scheme is based on the adiabatic
theorem~\cite{Messiah61b}.  Ideally the change in the control parameters
should be infinitely slow.  At a finite rate of change, there can be a
transition out of the computational subspace.  However, typically such a
Landau-Zener-type transition occurs with an exponentially small
probability $P\sim\exp(-\pi\Delta/\eta)$, where $\hbar\Delta$ is the
smallest energy gap between the computational subspace and the nearby
energy levels and $\eta$ is the adiabaticity parameter (i.e.,
\begin{math}
\frac{\partial}{\partial t}H(t) \sim \eta\hbar\Delta
\end{math})~\cite{ZenerComment,Berry90a,HwangJT77a,Dykhne61a,Zener32a}.
In Josephson networks the level distance is of order of the Josephson
coupling energy,
\begin{math}
\Delta \sim E_J/\hbar \sim (80\,\mathrm{ps})^{-1}
\end{math}.
For an operation time
\begin{math}
\tau_\mathrm{op} =1/\eta \gtrsim 3/\Delta
\end{math},
$P\lesssim10^{-4}$.  (For comparison, in a recent experiment concerning
dynamic quantum computation on Josephson qubits~\cite{Nakamura99a}, the
switching time was no shorter than
\begin{math}
0.25\hbar/E_J \sim 20\,\mathrm{ps}
\end{math}.)

In the previous section, on each superconducting box we neglected higher
charge states other than the two lowest.  In dynamic quantum computation
schemes, the existence of those higher levels may cause the quantum
leakage errors, i.e., it leads to a nonzero probability of leakage out
of the computational space, and more severely to renormalization of the
energy levels in the computational space (which therefore reduces the
gate fidelity)~\cite{Fazio99b}.  In Josephson qubits, however, the
coupling to the higher charge states are only through the Josephson
tunneling of Cooper pairs, which can be easily included in our
considerations [see Eqs.~(\ref{Holonomic:Prototype}),
(\ref{Holonomic:H3}), (\ref{Holonomic:H4}), and (\ref{Holonomic:H5})].
Those higher charge states form well-separated energy levels, and do not
alter the degenerate structure of the subspace in question, at least up
to the order of $(E_J/E_C)^2$ [in the dynamic scheme the quantum leakage
error occurs already in the order of $(E_J/E_C)^2$, see
Ref.~\onlinecite{Fazio99b}].  The leakage to the higher charge states
out of the computational subspace can therefore be considered within the
framework of Landau-Zener tunneling, which has already been discussed
above.  The renormalization of (degenerate) energy of the computational
space is irrelevant in our geometric scheme since it does not rely on
the dynamical time-evolution operator but only on the purely geometric
means.

In reality there are fluctuations of the (reduced) flux $\phi$ (tuning
the junctions) and the gate-induced charge $2en_g$ (resulting from the
fluctuations of random charges in the substrate or gate voltage itself).
One consequence of these fluctuations is the Landau-Zener-type
transitions out of the computational subspace.  A recent experiment on
Josephson charge qubits~\cite{Nakamura01z} suggests that fluctuations of
$n_g$ as well as $\phi$ are dominated by low frequency fluctuations.
Therefore, the Landau-Zener-type transitions might be small.
The fluctuations of $n_g$ can cause another type of errors: While the
eigenspace is by construction robust against the low-frequency
fluctuations of $\phi$, the random charge fluctuations lift the
degeneracy of the computational subspace.  The wave function of the
system then acquires dynamically accumulated phase factors
\begin{math}
\exp(-i\delta{E}\tau_\mathrm{op}/\hbar)
\end{math},
where $\delta{E}$ is the small level spacing caused by the fluctuations
of $n_g$.  Such dynamical phases can be ignored for sufficiently small
fluctuations and sufficiently short -- yet long
enough for adiabaticity -- operation time (%
\begin{math}
1/\Delta\ll\tau_\mathrm{op}\ll1/\delta{E}
\end{math}).

Another common source of decoherence in Josephson charge qubits is the
quasi-particle tunneling~\cite{Schon90a}.  In particular, since the
computational eigenspace is not the lowest energy states
[Eq.~(\ref{Holonomic:Prototype}) and Fig.~\ref{Holonomic:fig2} (b)], it
gives rise to the relaxation out of the eigenspace to lower energy
states (this effect cannot be described by the Landau-Zener-type
transitions).  At sufficiently low temperatures compared with the
superconducting gap $\Delta_S$, the quasi-particle tunneling rate
$\Gamma_\mathrm{qp}$ is exponentially small~\cite{Schon90a},
\begin{math}
\Gamma_\mathrm{qp} \sim \exp[-(2\Delta_S+E_C)/k_BT]
\end{math}.
For example, in the experiment on Cooper-pair box~\cite{Nakamura99a},
\begin{math}
\Gamma_\mathrm{qp} \sim (6\, \mathrm{ns})^{-1} \sim 10^{-2}E_J/\hbar
\end{math}
at temperature $30\,\mathrm{mK}$ even through the probe junction which
was biased by a voltage
\begin{math}
eV \sim 2\Delta_S + E_C
\end{math}
(without the voltage bias $\Gamma_\mathrm{qp}$ should be even smaller).
In such a situation, one can therefore conclude that the effect of
quasi-particles is negligible.

For a brief comparison of the present scheme with the conventional
Josephson charge qubit~\cite{Makhlin99a,Nakamura99a}, we estimate the
fidelity for a single phase-shift operation $U_Z$.  The fidelity in this
case is given by
\begin{equation}
\label{paper::eq:fidelity}
\mathrm{fidelity} \simeq \sqrt{(1-P)[1-\sin^4(\delta\phi/2)]} \,,
\end{equation}
where $P$ is the probability for Lanau-Zener-type transitions or
quasiparticle tunnelings to occur and $\delta\phi$ is the error in phase
shift due to the background charge fluctuations, i.e., $\delta\phi =
\delta{E}\tau_\mathrm{op}/\hbar$ (see above).  Taking
\begin{math}
\Delta = (80\, \mathrm{ps})^{-1}
\end{math},
$\tau_\mathrm{op}=3/\Delta$, $\Delta_S = 5\Delta$, $k_BT = \Delta/10$,
and $\delta{E}/\hbar = \Delta/10$ (see Ref.~\onlinecite{Nakamura99a}),
one estimates
\begin{math}
\mathrm{fidelity} \simeq 0.998
\end{math}.
In the dynamic scheme the fidelity is also given by the same form as
Eq.~(\ref{paper::eq:fidelity}).  The differences are that $P$ is mainly
responsible for the quasiparticle tunnelings and that the phase error
$\delta\phi$ comes from the finite ramping time of the gate pulse.
Taking the parameters from the recent experiment~\cite{Nakamura99a}, we
see that the fidelity takes the same value (up to three digits below
decimal point).

Lastly, in ideal case some tunable junctions [e.g., $J_1$ in
Eq.~(\ref{Holonomic:H3}), see discussions below
Eq.~(\ref{Holonomic:Ket1b})] need to be turned off completely.  In
reality, the Josephson energies of the two parallel junctions on a SQUID
loop (Fig.~\ref{Holonomic:fig1}) may not be identical [i.e.,
$\gamma\neq1$ in Eq.~(\ref{Holonomic:Junction}) and
Fig.~\ref{Holonomic:fig1}].  Then, a tunable junction (i.e.  SQUID)
cannot be turned off completely~\cite{Falci00a}.  This makes it
nontrivial to prepare an initial state which should be a product state
of the ``true'' qubit and the ancilla qubits in a logical block [see,
e.g., $\ket{\lambda_1}$ below Eq.~\eqref{Holonomic:H3}].  In practice,
such a difficulty can be overcome by means of fast relaxation processes
with the gate voltages of the ancilla qubits adjusted far off the
resonance in the initial state-preparation stage.  This process also
allows for preparation of the ``true'' qubit in a definite initial
state~\cite{Makhlin99a}.

\section{Conclusion}

We have proposed a scheme based on geometric means to implement quantum
computation on solid-state devices.  The main advantage of a geometric
computation scheme is its intrinsically fault-tolerant
feature~\cite{Preskill99a,Ellinas01a}.  However, it is usually
non-trivial to find a physical system whose Hamiltonian has a particular
degenerate structure for geometric computation.  The scheme discussed in
this work provides a generic way to construct such a system from
arbitrary quantum two-state systems as long as couplings satisfy certain
requirements specified above.  Such requirements are rather easy to
fulfill on solid-state devices.  A drawback of this scheme is that it
requires more resources (four Cooper-pair boxes per each qubit).
Considering quantum error correcting codes, however, it may not be a
major disadvantage.  Moreover, since the current scheme is based on
adiabatic evolution, it does not require sharp pulses of flux and gate
voltage.  With current technology, it is still challenging to obtain a
sufficiently sharp pulses of flux and gate voltages (in
Ref.~\onlinecite{Nakamura99a}, the raising and falling times were of
order of $\hbar/E_J$).  Finite raising and falling times of pulses can
result in a significant error in dynamic computation
schemes~\cite{ChoiMS01a}.

\bigskip
\begin{acknowledgments}
The author thanks J.-H.~Cho, C.~Bruder, I.~Cirac, R.~Fazio, and
J.~Pachos for discussions and comments.  This work was supported by a
Korea Research Foundation Grant (KRF-2002-070-C00029).
\end{acknowledgments}


\end{document}